\documentstyle[12pt]{article}
\begin{document} \begin{center}

{\bf SURFACE DENSITIES IN GENERAL RELATIVITY\\}
 \vspace{1.5cm} \small{L. FERN\'ANDEZ-JAMBRINA} and \small{F. J. CHINEA}\\ {\it
Departamento de F\'{\i}sica Te\'orica II, Facultad de Ciencias F\'{\i}sicas
\\Ciudad Universitaria, E-28040 Madrid, Spain\\}
 \vspace{1cm} \small{ABSTRACT}\\  

 \baselineskip 12pt \parbox{30pc}{\vspace{2mm}\footnotesize{\hspace{.7pc} 
In this lecture we deal with the construction of surface densities for the
angular momentum of the sources of asymptotically flat vacuum stationary
axisymmetric spacetimes. These sources arise from the discontinuities of the
twist potential. The result will be applied to the Kerr metric to obtain an
integrable density which can be viewed as the regularized version of the density
obtained using other formalisms.}} \end{center}

\baselineskip 14pt 
\vspace{1cm} 
\noindent{\bf 1.Introduction}
\vspace{.5cm}

The construction of compact inner sources for stationary vacuum solutions of the
Einstein equations is one of the challenges of general relativity. So far nobody
has been able to match a physically reasonable matter spacetime to a nonstatic
vacuum exterior. In this lecture we shall be less ambitious and restrict
ourselves to shells that could act as minimal sources and calculate the surface
densities for dipolar physical quantities such as the angular momentum. There
is already a previous formalism due to Israel [1] which is grounded on the thin
layer theory and Lanczos jump conditions [2]. Here we shall develop a different
approach which is closer to the classical potential theory [3].

In order to accomplish this task it will be shown in section 2 a classical
calculation of the magnetic moment density, which will be useful as a toymodel
for subsequent relativistic results. This will be generalized for the angular
momentum in general relativity in section 3 and the result will be applied to
the Kerr metric [4] in section 4 [5]. A discussion of the results is provided at
the end of this essay.

 \vspace{1cm}
\noindent{\bf 2.Classical dipolar surface densities}
 \vspace{.5cm}

In this section it is included a construction for obtaining classical
densities of dipole moment from the discontinuities of the magnetic scalar
potential. This result is consistent with the classical potential theory [3] and
allows the calculation of sheets of magnetic dipoles as sources for the
magnetostatic field. 

For this purpose we need review both descriptions of the magnetostatic field in
vacuum:

\begin{equation}
{\bf B}=-\nabla V=\nabla\times {\bf A}
\end{equation}

It is well known that the field can always be written as the rotor of the
potential vector ${\bf A}$, since the divergence of ${\bf B}$ is zero. On the
other hand the field can only be expressed as the gradient of a scalar
potential $V$ out of the sources and if the magnetic field is static. If the
topology of the source happened not to be trivial, (think, for instance, of a
ring of electric current) the scalar potential could not be defined as a
continuous function in the vacuum surrounding the source.

The following formulae from integral calculus will also de useful:

\begin{equation}
\int_\Omega d^3x\nabla V=\int_{\partial\Omega} dS\,{\bf n}\,V
\end{equation}

\begin{equation}
\int_\Omega d^3x\nabla\times{\bf A}=\int_{\partial\Omega}dS\,{\bf n}\times{\bf
A} 
\end{equation}

These expressions relate integrals in a volume $\Omega$ with integrals on its
boundary. Although these formulae resemble much the Stokes theorem, they are
not a direct consequence of it, since the Green identities (which are
metric-dependent) are needed in their derivation. Therefore they rely much
on the fact that the spacetime is flat and it is not straightforward to
generalize them.

Let us assume that the magnetic field behaves at great distances from
its source as the one produced by a magnetic moment $M$ in the direction of $z$.
This means that the potentials can be written as:

\begin{equation}
V=\frac{M\,\cos\theta}{r^2}+0(r^{-3})\ \ \ \ \ A=\frac{M\,\sin\theta}{r^2}\,{\bf
u_z}+0(r^{-3}) 
\end{equation}

The integral of the difference of both expressions for ${\bf B}$ will be
obviously zero:

\begin{equation}
0=\int_{{\bf R}^3}d^3x(\nabla\times{\bf A}+\nabla V)=
 \int_{\partial{\bf R}^3}dS({\bf n}\times{\bf A}+{\bf n} V)
\end{equation}

Let us assume that the scalar potential is discontinuous across a closed surface
$S$: The integral must be split into two pieces at $S$ and this surface has to
be taken as a part of the boundary, as well as the sphere at infinity.

The integral at infinity is straightforward to calculate since the asymptotic
values of the potentials are known. We are left then with the following
integral:

\begin{equation}
M=\frac{1}{4\pi}\,\int_S\,dS\,[V]\,{\bf n}\cdot{\bf u_z}\label{clasmag}
\end{equation}
where $[V]$ denotes the jump of the scalar potential and ${\bf n}$ is the
outward unitary normal to the surface.

According to potential theory we can interpret the integrand as the
magnetic moment surface density of the source:

\begin{equation}
\sigma_M=\frac{1}{4\pi}\,[V]\,{\bf n}\cdot{\bf u_z}\label{classic}
\end{equation}

\newpage

 \noindent{\bf 3.Relativistic dipole densities}
\vspace{.5cm}

In order to generalize the previous expression to general relativity and apply it
to angular momentum [5], we shall make use of some results included in [6] to
obtain two different expressions for the rotation vector $\omega$ of the
congruence of worldlines of constant spatial coordinates (that is, the dragging
of inertials):

\begin{equation}
\omega=-f^{-1}\,d\chi=-\rho^{-1}\,f\,*dA
\end{equation}
where $\chi$ is the twist potential [7] , $*$ stands for the Hodge dual in
the space orthogonal to the orbits of the Killing vectors and the other functions
can be read from the general expression for a stationary axisymmetric vacuum in
Weyl coordinates:

\begin{eqnarray}
ds^{2}=-f\,(dt-A\,d\phi)^{2}+f^{-1}[e^{2\,k}(d\rho^{2}+dz^{2})+\rho^2
d\phi^{2}]\label{eq:can}
\end{eqnarray}

Asymptotic flatness will be imposed on the metric functions and the twist
potential in terms of the mass, $m$, and angular momentum, $J$, of the source
in an adequate system of coordinates:

\begin{equation}
f=1-\frac{2\,m}{r}+O(r^{-2})\ \ \ \ \ k=O(r^{-2})
\end{equation}

\begin{equation}
A=-\frac{2\,J\sin^2\theta}{r}+O(r^{-2})\ \ \ \ \ 
\chi=-\frac{2\,J\cos\theta}{r^2}+O(r^{-3})
\end{equation}

As we did in the previous section, we can calculate the following integral over
the metric space $(V_3,^3g)$ orthogonal to the hipersurfaces of constant time:

\begin{equation}
0=\int_{V_3}\sqrt{^3g}\,f^{-1/2}\,<f^{-1}\,d\chi-\rho^{-1}\,f\,*dA,dZ>\,dx^1dx^2dx^3
\end{equation}

The function $Z$ involved in the scalar product is defined by the following
elliptic differential equation:

\begin{eqnarray}
\partial_{\mu}(\sqrt{^3g}\,f^{-3/2}g^{\mu\nu}\partial_{\nu}Z)=0\label{eq:lap}
\end{eqnarray}
with boundary condition $Z\sim r\,\cos\theta$ at infinity.

Under these conditions the previous integral can be reduced to a surface
integral over the boundary of $V_3$, that will be again the sphere at infinity
and a surface $S$ where $\chi$ is assumed to be discontinuous. The final
expression is rather similar to the one obtained previously for the magnetic
moment:

\begin{equation}
J=\int_S\,dS\,\sigma_J\ \ \ \ \
\
\sigma_J=-\frac{1}{8\,\pi}[\chi]\,f^{-3/2}\,g^{\mu\nu}\,n_\mu\,\partial_{\nu}Z
\label{eq:den}
 \end{equation}
where we denote by $[\chi]$ the jump of the twist potential across $S$.

 \noindent{\bf 4.An example: The Kerr metric}
\vspace{.5cm}

Finally we shall apply the previous result to a solution of astrophysical
interest, the Kerr metric:

\begin{eqnarray}ds^2&=&-
(1-\frac{2mr}{r^2+a^2\cos^2\theta})(dt+\frac{2mar\sin^2\theta}{r^2-2mr+a^2\cos^2\theta}
d\phi)^2+\nonumber\\&+&(1-\frac{2mr}{r^2+a^2\cos^2\theta})^{-1}\{(r^2-2mr+a^2)\sin^2\theta
d\phi^2
+\nonumber\\&+&(r^2-2mr+a^2\cos^2\theta)(
\frac{dr^2}{r^2-2mr+a^2}+d\theta^2)\} 
\end{eqnarray}
whose twist potential is given by the following expression:

\begin{equation}
\chi=-\frac{2\,m\,a\,\cos\theta}{r^2+a^2\,\cos^2\theta}
\end{equation}

This potential can be shown to be discontinuous [5] across
the disk $r=0$,  $\theta\in[0,\pi/2)$ if we identify events with collatitude
$\theta$ with those with $\pi-\theta$ on the disk, as it is done in [8] to avoid
the inclusion of regions with negative radius:

\begin{equation}
[\chi]_{r=0}=-\frac{4\,m}{a\cos\theta}
\end{equation}

The disk is flat and its surface element and unitary normal are given by the
following expressions:

\begin{eqnarray}
ds^2=a^2(\cos^2\theta d\theta^2+\sin^2\theta d\phi^2)\ \ \ \ \ 
n=\frac{1}{\cos\theta}\partial_r
\end{eqnarray}

Therefore we only need the required function $Z$ to calculate the angular
momentum surface density of the source hidden in $S$. A solution for
(\ref{eq:lap}) satisfying the boundary condition at infinity is:

\begin{eqnarray}
Z=(r-3m)\cos\theta+\frac{2\,a^2\,m\,\cos^3\theta}{r^2+a^2\,\cos^2\theta}
\end{eqnarray} 

The surface density constructed for the Kerr metric according to
(\ref{eq:den}) is then:

\begin{eqnarray}
\sigma_J=\frac{m}{2\pi a\cos\theta}\label{eq:reg}
\end{eqnarray} 

And if we integrate it over the disk $r=0$, it yields the correct result for
the angular momentum of the Kerr metric:

\begin{equation}
J=m\,a
\end{equation}

\vspace{1cm}
 \noindent{\bf 5.Discussion} 
\vspace{.5cm}

If we compare the angular momentum density for the disk $r=0$ obtained with
this new formalism with the one calculated by Israel in [8], we notice that they
are somewhat different. In that reference, the surface density is non-integrable:

\begin{equation}
\sigma=-\frac{m\,\sin^2\theta}{4\,\pi\,a\,\cos^3\theta}
\end{equation}
 and the singular ring has to be
included [9] to correct the result. Therefore, what we have calculated is a
regularized version of Israel's result.

This is not unusual in potential theory: Think for instance of the magnetic
field generated by a ring of constant electrical current. If the formula
(\ref{classic}) is used, one obtains a uniform sheet of magnetic dipoles
instead.

The extension of these results is discussed in
references [10].

 \vspace{1cm}
 \noindent{\bf 6.Acknowledgements} 
\vspace{.5cm}

The present work has been supported in part by DGICYT Project PB92-0183. L.F.J. 
is supported  by a FPI Predoctoral Scholarship from Ministerio de Educaci\'{o}n y
Ciencia (Spain). The authors wish to thank L. M. Gonz\'alez-Romero
and J. A. Ruiz-Mart\'\i n for discussions.

\vspace{1cm}
 \noindent{\bf 7.References} 
\vspace{.5cm}

\noindent 1. W. Israel, {\it Nuovo Cimento}\/ {\bf 44 B}, 1 (1966).\\
\noindent 2. C. Lanczos, {\it Ann. Physik}\/ {\bf 74}, 518 (1924).\\
\noindent 3. O. D. Kellogg, {\it Foundations of Potential Theory}\/ Dover,
New York, (1954).\\
\noindent 4. R. P. Kerr, {\it Phys. Rev. Lett.} \/{\bf 11}, 1175 (1963).\\
\noindent 5. L. Fern\'andez-Jambrina and F. J. Chinea, {\it Phys. Rev.
Lett.}\/ {\bf 71}, 2521 (1993) [arXiv: gr-qc/0403102]\\
\noindent 6. F. J. Chinea and L. M. Gonz\'alez-Romero, {\it Class. Quantum
Grav.}\/ {\bf 9}, 1271 (1992).\\
\noindent 7. F. J. Ernst, {\it Phys. Rev.}\/ {\bf 167}, 1175 (1968).\\
\noindent 8. W. Israel, {\it Phys. Rev.}\/ {\bf D 2}, 641 (1970).\\
\noindent 9. C. L\'opez, {\it Nuovo Cimento}\/ {\bf 66 B}, 17 (1981).\\
\noindent 10. L. Fern\'andez-Jambrina [arXiv: 
gr-qc/0403113, arXiv: gr-qc/0404008], 
L. Fern\'andez-Jambrina  and F. J. Chinea [arXiv: gr-qc/0403118]

  \end{document}